\documentclass[letter]{aa} % for the letters 
\usepackage{graphicx}
\usepackage{txfonts}
\usepackage{graphicx,graphics,rotating,amssymb,amsmath}
\usepackage{xcolor}
\usepackage{multirow}
\usepackage{longtable}
\usepackage{booktabs}
\usepackage{subfig}
\usepackage{float,capt-of}
\usepackage{natbib}
\usepackage[english]{babel}
\usepackage{morefloats}
\usepackage{color}
\usepackage{sidecap}
\usepackage{epsfig,color}
\usepackage{wrapfig}
 \usepackage{lscape}
\usepackage{url}
\newcommand\kms{km~s$^{-1}$}

\begin{document} 

\title{SUPER. VI. A giant molecular halo around a $z\sim2$ quasar}
%\subtitle{I. Overviewing the $\kappa$-mechanism}

\author{C. Cicone \thanks{claudia.cicone@astro.uio.no}
                \inst{1}
                \and
        V. Mainieri\inst{2} 
                \and
               C. Circosta\inst{3}
            \and 
            D. Kakkad\inst{4,16}
            \and
            G. Vietri\inst{17}
            \and 
            M. Perna\inst{5}
            \and 
            M. Bischetti\inst{6}
            \and 
            S. Carniani\inst{7}
            \and 
           G. Cresci\inst{8}
           \and
         C. Harrison\inst{9}
         \and
         F. Mannucci\inst{8}
         \and 
         A. Marconi\inst{8,10}
         \and
         E. Piconcelli\inst{11}
         \and
         A. Puglisi\inst{12}
         \and 
         J. Scholtz\inst{13}
         \and 
         C. Vignali\inst{14, 15}
     \and
     G. Zamorani\inst{15}
 \and
 L. Zappacosta\inst{11}
 \and 
 F. Arrigoni Battaia\inst{18}}

\institute{Institute of Theoretical Astrophysics, University of Oslo, P.O. Box 1029, Blindern, 0315 Oslo, Norway %1
        \and European Southern Observatory, Karl-Schwarzschild-Str. 2, 85748
Garching bei M\"{u}nchen, Germany %2
        \and Department of Physics and Astronomy, University College London, London WC1E 6BT, UK %3
        \and European Southern Observatory, Alonso de Cordova 3107, Casilla
19, 19001 Santiago, Chile %4
\and Centro de Astrobiologi{\'i}a (CSIC-INTA), Ctra. de Ajalvir, Km 4, 28850, Torrej{\'o}n de Ardoz, Madrid, Spain %5
\and INAF - Osservatorio Astronomico di Trieste, Via G. B. Tiepolo 11, I–34143 Trieste, Italy %6
\and Scuola Normale Superiore, Piazza dei Cavalieri 7, 56126 Pisa, Italy %7
\and INAF - Osservatorio Astrofisico di Arcetri, Largo E. Fermi 5,  50125, Firenze, Italy %8
\and School of Mathematics, Statistics and Physics, Newcastle University, Newcastle upon Tyne, NE1 7RU, UK %9 
        \and Dipartimento di Fisica e Astronomia, Università di Firenze, Via G.
Sansone 1, 50019 Sesto Fiorentino (Firenze), Italy  %10
        \and INAF – Osservatorio Astronomico di Roma, Via Frascati 33, 00040 Monte Porzio Catone, Roma, Italy %11
        \and Centre for Extragalactic Astronomy, Durham University, South Road, Durham DH1 3LE, UK %12
\and Chalmers University of Technology, Department of Space, Earth and Environment, Onsala Space Observatory, 439 92 Onsala, Sweden %13
\and Dipartimento di Fisica e Astronomia, Università degli Studi di
Bologna, via P. Gobetti 93/2, 40129 Bologna, Italy %14
\and INAF/OAS, Osservatorio di Astrofisica e Scienza dello Spazio di
Bologna, via P. Gobetti 93/3, 40129 Bologna, Italy %15
\and Department of Physics, University of Oxford, Denys Wilkinson Building, Keble Road, Oxford, OX1 3RH, UK %16
\and INAF – Istituto di Astrofisica Spaziale e Fisica Cosmica di Milano, Via A. Corti 12, I-20133 Milano, Italy %17
\and Max-Planck-Institute f\"{u}r Astrophysik, Karl-Schwarzschild-Strasse 1, D-85748 Garching bei M\"{u}nchen, Germany %18
}

\date{Received 22 Jun 2021 / Accepted: 02 Sep 2021 }

%________________________________________________________________
% ABSTRACT
%________________________________________________________________

\abstract{
We present the discovery of copious molecular gas in the halo of cid\_346, a $z=2.2$ quasar studied as part of the SINFONI survey for Unveiling the Physics and Effect of Radiative feedback (SUPER). New Atacama Compact Array (ACA) CO(3-2) observations detect a much higher flux (by a factor of $14\pm5$) than measured on kiloparsec scales ($r\lesssim8$~kpc) using previous snapshot Atacama Large Millimeter/submillimeter Array data. Such additional CO(3-2) emission traces a structure that extends out to $r\sim200$~kpc in projected size, as inferred through direct imaging and confirmed by an analysis of the {\it uv} visibilities. This is the most extended molecular circumgalactic medium (CGM) reservoir that has ever been mapped. It shows complex kinematics, with an overall broad line profile (FWHM $= 1000$~\kms) that is skewed towards redshifted velocities up to at least $v\sim1000$~\kms.
Using the optically thin assumption, we estimate a strict lower limit for the total molecular CGM mass observed by ACA of $M_{mol}^{CGM}>10^{10}~M_{\odot}$. There is however room for up to $M^{CGM}_{mol}\sim 1.7\times 10^{12}$~$M_{\odot}$, once optically thick CO emission with $\alpha_{\rm CO}=3.6$~$\rm M_{\odot}~(K~km~s^{-1}~pc^2)^{-1}$ and $L^{\prime}_{CO(3-2)}/L^{\prime}_{CO(1-0)}=0.5$ are assumed. 
Since cid\_346 hosts quasar-driven ionised outflows and since there is no evidence of merging companions or an overdensity, we suggest that outflows may have played a crucial rule in seeding metal-enriched, dense gas on halo scales. However, the origin of such an extended molecular CGM remains unclear.}

\keywords{galaxies:active; (galaxies:)quasars:individual; (galaxies:) intergalactic medium; galaxies:halos; galaxies:high redshift; submillimeter: galaxies}

\maketitle

%-------------------------------------------------------------------

\section{Introduction}\label{sec:introduction}

The past few years of observational and theoretical advances in the field of galaxy formation and evolution have drawn significant attention to the role of gaseous halos surrounding galaxies, referred to as `circumgalactic medium' (CGM, usually defined to extend up to the virial radius). The CGM is continuously enriched by galactic outflows, cosmological inflows, and mergers, and it is depleted by the same processes. The CGM gas undergoes several physical and chemical transformations, resulting in a constantly evolving multi-phase medium. 
Originally expected to be in a rarefied, highly ionised diffuse phase at $T\sim10^6~K$ (the `Galactic corona'), the CGM has been studied mostly in absorption \citep{Tumlinson+17,Chen17_chap9}. However, there is now overwhelming evidence that the CGM of both normal and active galaxies, at any redshift, also embeds much colder and denser gas clouds, which allow it to be detected in emission using several tracers: Ly$\alpha$ \citep{ArrigoniBattaia+19, Cai+19, Bacon+21}, CIV and HeII \citep{Travascio+20, Guo+20}, [OII] \citep{Rupke+19}, MgII \citep{Burchett+21}, and H$\alpha$ \citep{Fossati+19}. Nevertheless, the total mass of (multi-phase) gas stored on CGM scales remains unconstrained.

In addition to (tidal or ram-pressure) gas stripping from satellites (e.g. \citealt{Nelson+20}), galactic outflows and fountains are the primary mechanisms enriching the CGM with metals and cool gas \citep{Suresh+19}. Observations have shown that outflows can carry large amounts (a few $\sim 10^{10}~M_{\odot}$) of cold and dense molecular (H$_2$) gas travelling at speeds $v>1000$~\kms, out to $r\sim10$~kpc (see review by \citealt{Veilleux+20}). Although there is not yet evidence for a significant H$_2$ gas mass in the CGM, \cite{Cicone+19} have pointed out that these observations will only become feasible with a new, large (e.g. 50-m) single dish (sub)mm telescope because of the limited sensitivity of current facilities to diffuse large-scale structures of cold gas. These limitations are partially mitigated at high-$z$, 
thanks to the larger angular diameter distance, and indeed there have been a few promising detections at $z>2$. The [C {\sc ii}]~$158\mu m$ line, probing both H{\sc i} and H$_2$ gas, was imaged out to $r\sim30$~kpc around a luminous $z=6.4$ quasar hosting a massive outflow \citep{Cicone+15} and out to $r\sim10$~kpc in a number of individual main sequence galaxies at $4<z<6$ \citep{Fujimoto+20}. Diffuse CO and [C{\sc i}] components have been revealed on scales of $\sim70$~kpc around the Spiderweb galaxy \citep{Emonts+16} and $\sim40$~kpc around a massive star forming galaxy at $z=3.5$ \citep{Ginolfi+17}. 

The target of this study, cid\_346, is an X-ray selected $z\sim2$ type 1 active galactic nucleus (AGN), which is part of the SINFONI survey for Unveiling the Physics and Effect of Radiative feedback (SUPER\footnote{\textrm{http://www.super-survey.org}}, \citealt{Circosta+18}). The AGN ($L_{Bol, AGN}=10^{46.66}~erg~s^{-1}$) and its host galaxy (SFR$=360$~M$_{\odot}~yr^{-1}$, $M_*=10^{11}~M_{\odot}$, see target properties in Table~\ref{table:properties}) are likely undergoing an explosive feedback phase since energetic outflows have been detected from parsec ($v_{\rm CIV}^{wind}\sim2230$~km~s$^{-1}$, \citealt{Vietri+20}\footnote{Measured from the broad, blue-shifted CIV emission component. The spectrum also exhibits a CIV broad absorption line (BAL).}) to kiloparsec scales ($v_{\rm [OIII]}^{out}=1700$~km~s$^{-1}$, \citealt{Kakkad+20}). Using snapshot $1~\arcsec$-resolution Atacama Large Millimeter/submillimeter Array (ALMA) observations, \cite{Circosta+21} resolved the CO(3-2) stemming from the interstellar medium (ISM), and found no evidence for depletion of H$_2$ gas with respect to inactive galaxies at matched $z$, $M_{*}$, and broadly covering the same  star formation rate (SFR) range. In this Letter we present new Atacama Compact Array (ACA) CO(3-2) observations of cid\_346 tracing significant additional CO emission beyond the ISM, which led to the first image of a molecular halo out to $r\sim200$~kpc.

\section{Observations}\label{sec:obs}

The technical parameters, including line sensitivities, of the data used in this work are reported in Table~\ref{table:obs}. In the following, we briefly outline the data reduction and analysis steps.

The ACA CO(3-2) line observations are dated 6-8 March 2020 (Project code 2019.2.00118.S, PI: Mainieri). Data reduction and analysis were performed using the Common Astronomy Software Applications package (CASA, \citealt{McMullin+07}). We obtained the calibrated measurement set by running the standard pipeline through the {\it scriptForPI} in CASA v.5.6. The continuum is not detected by ACA (see Appendix~\ref{sec:continuum}), hence we did not perform a continuum subtraction.
The data were imaged at their native spectral resolution of $\Delta v_{\rm channel} =5.45$~\kms.  Deconvolution and cleaning were executed using the task \texttt{tclean} with interactive guidance for mask selection. We applied Briggs weighting and set the robust parameter equal to 0.5. All spectra were extracted from the primary-beam corrected cubes.

The ALMA CO(3-2) data are the same as in \cite{Circosta+21}. We reduced them using the CASA pipeline (v.5.1) and subtracted the continuum in the {\it uv} 
plane using \texttt{uvcontsub} with a linear polynomial fit, estimating the continuum from the $\nu_{obs}<107.2$~GHz and $\nu_{obs}>107.6$~GHz channels, corresponding to $|v|>560$~\kms.
We note that this dataset spans $\nu_{obs}\in(107.07, 109.5)$~GHz, corresponding to $v\in(-5900, 920)$~km~s$^{-1}$, hence it does not provide good coverage of the redshifted side of the CO line. The maximum spectral resolution is 22~km~s$^{-1}$. We imaged the ALMA {\it uv} visibilities with Briggs weighting and robust=0.5. We also produced a second, lower resolution cube by applying an {\it uv} tapering of $4\arcsec$ to enhance any extended components (ALMA-t in Table~\ref{table:obs}).

The APEX $^{12}$[CI]$^3$P$_{2}$-$^3$P$_{1}$ (hereafter, [CI](2-1)) line observations were carried out in 2017 (five UT dates in April-June 2017, 
project ID: 098.A-0774, PI: Cicone) and 2018 (4 UT dates in May 2018, 
project ID: 0101.B-0758, PI: Cicone) with the PI230 heterodyne receiver in dual polarisation. Jupiter and IRC+10216 were the main focus and pointing calibrators.
We placed the tuning frequency ($\rm \nu_{obs}=251.372$~GHz) at IF=10 GHz in the upper side band (USB), and analysed the data exploiting the full IF 4-12 GHz bandwidth. The instrumental resolution is 0.061~MHz ($\Delta v_{res}$= 0.073~\kms).  
We reduced the data using the Continuum and Line Analysis Single-dish Software (CLASS), which is part of the GILDAS package\footnote{\texttt{http://www.iram.fr/IRAMFR/GILDAS}}. We checked all scans and dropped those showing baseline ripples or instrumental features. The final spectrum (Figure~\ref{fig:APEX_CI_spec}) does not show any clear [CI] detection, down to a 1$\sigma$ rms line sensitivity of $T_{A}^{\prime}=0.175$~mK in $\Delta v=50$~\kms. To convert the antenna temperature units into flux density, we adopted a factor of $40\pm7$ Jy~K$^{-1}$, which is the average value estimated for PI230 during our observing runs\footnote{\texttt{http://www.apex-telescope.org/telescope/efficiency/}}.

\begin{table}[tbp]
        \centering \small
        \caption{Source properties: cid\_346}
        \label{table:properties}
        \begin{tabular}{lc}
                \hline
                \hline
                 RA, Dec (ICRS) & 09:59:43.412, 02:07:07:402 \\
                 Redshift & $z_{\rm CO}=2.2197^{\S}$ \\
                 Physical scale [kpc~arcsec$^{-1}$] & 8.459~$^{*}$ \\
                Nuclear activity & quasar \\
                 $\log M_{*}~[M_{\odot}]$ &  $11.0\pm0.2$$^{\dag}$ \\
                 SFR [$M_{\odot}~yr^{-1}$]  & $360\pm 50$$^{\dag}$ \\
                 $\log L_{\rm Bol, AGN}~[\rm erg~s^{-1}]$ & $46.66\pm 0.02$$^{\ddag}$ \\
                 $\log M_{BH}$ $[M_{\odot}]$ & $9.1\pm 0.3$$^{\ddag}$ \\
                 $\lambda_{Edd} = L_{\rm Bol, AGN}/L_{Edd}$ & $0.2$$^{\ddag}$ \\

                \hline
        \end{tabular}
        
        \begin{flushleft}
                {\it Notes:} 
                $^{\S}$ Line peak of the ALMA CO(3-2) spectrum probing $r\lesssim8$~kpc (see also \citealt{Circosta+21}).
                $^{*}$ We adopted a standard $\Lambda$ cold dark matter cosmology with $H_0$ = 67.4 \kms Mpc$^{-1}$, $\Omega_{\rm M}$ = 0.315, and $\Omega_{\Lambda}$ = 0.685 \citep{Planck2018_parameters}. $^{\dag}$ \cite{Circosta+18}; $^{\ddag}$ \cite{Vietri+20}.
        \end{flushleft}
\end{table}

\begin{table}[tbp]
        \centering \small
        \caption{Description of the sub-millimetre and millimetre data used in this work}
        \label{table:obs}
        \begin{tabular}{lccccc}
                \hline
                \hline
                \multicolumn{6}{c}{$^{12}$CO(J=3-2) line ($\nu_{\rm rest}=345.796$~GHz, $\nu_{obs} = 107.4$~GHz)} \\
                \hline 
                Tel. & Res.      & MRS/FoV &  Time & PWV        & 1$\sigma$ ($\Delta v_{channel})$  \\
                     & [$\arcsec$]  &   [$\arcsec$] &  [h]      & [mm]  & [mJy/beam] \\ 
                 (1) & (2)  & (3)   & (4)    & (5) & (6) \\ 
                \hline
                ACA &  $9.3$  & 73/99  &  5.2  & 5.2 & 4.0 (6~\kms)  \\
                ALMA & $1.0$ & \multirow{2}{*}{9.9/57} &  \multirow{2}{*}{0.2}  & \multirow{2}{*}{1.8}    & 0.7 (22~\kms) \\
                ALMA-t &  $3.4$  &       &     &   & 1.4 (22~\kms) \\
                \hline
                \hline
                \multicolumn{6}{c}{$^{12}$[CI]$^3$P$_{2}$-$^3$P$_{1}$ line ($\nu_{\rm rest}=809.342$~GHz, $\nu_{obs}=251.372$~GHz)}\\
                \hline 
                APEX &  27  & 27/27  &  16.9  & 1.5 & 7.0 (50~\kms)  \\
                \hline
        \end{tabular}
        \begin{flushleft}
                {\it Columns:} (1) telescope; (2) spatial resolution; (3) maximum recoverable scale (MRS) and field of View (FoV); (4) on source time; (4) precipitable water vapour (PWV); and (6) average line sensitivity per spectral channel ($1\sigma$ rms).
        \end{flushleft}
\end{table}

\begin{figure}[tb]
        \centering
        \includegraphics[clip=true,trim=2.5cm 7.9cm 2.cm 1.8cm,scale=.55,angle=0]{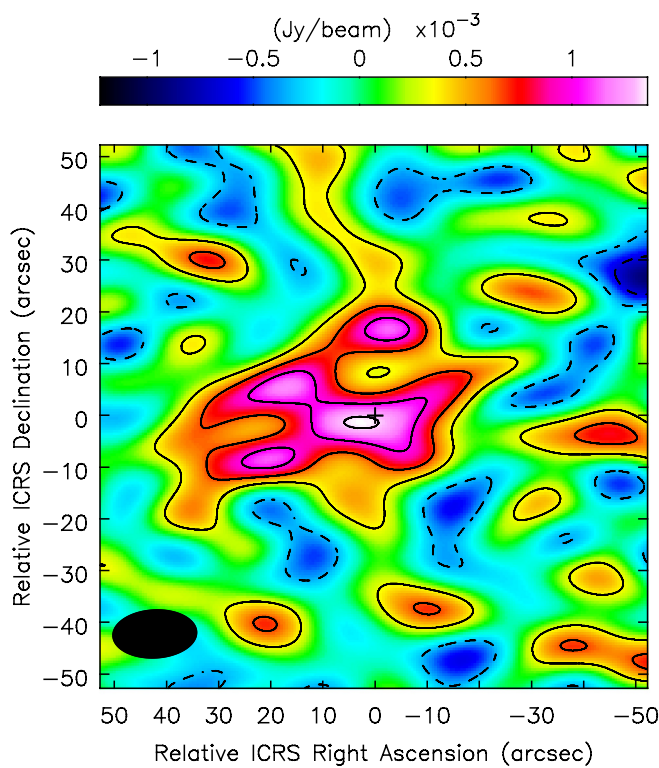}
        \caption{ACA map obtained by integrating the 
        		CO(3-2) line emission within 
                $-400<v[\rm km~s^{-1}]<1000$. For visualisation purposes, the map has not been divided by the primary beam profile, hence the noise is uniform across the field. Negative contours are shown as dashed curves. Both negative and positive contours are plotted in steps of $1\sigma$ starting from $\pm1\sigma$, with a $1\sigma=0.33$~mJy~beam$^{-1}$. The phase centre is indicated with a black cross (see coordinates in Table~\ref{table:properties}) and corresponds to the AGN position.
        }\label{fig:co32_aca_maps}
\end{figure}

\begin{figure}[tb]
        \centering
        \includegraphics[clip=true,trim=4.2cm 3cm 3.7cm 3.2cm,scale=.44,angle=270]{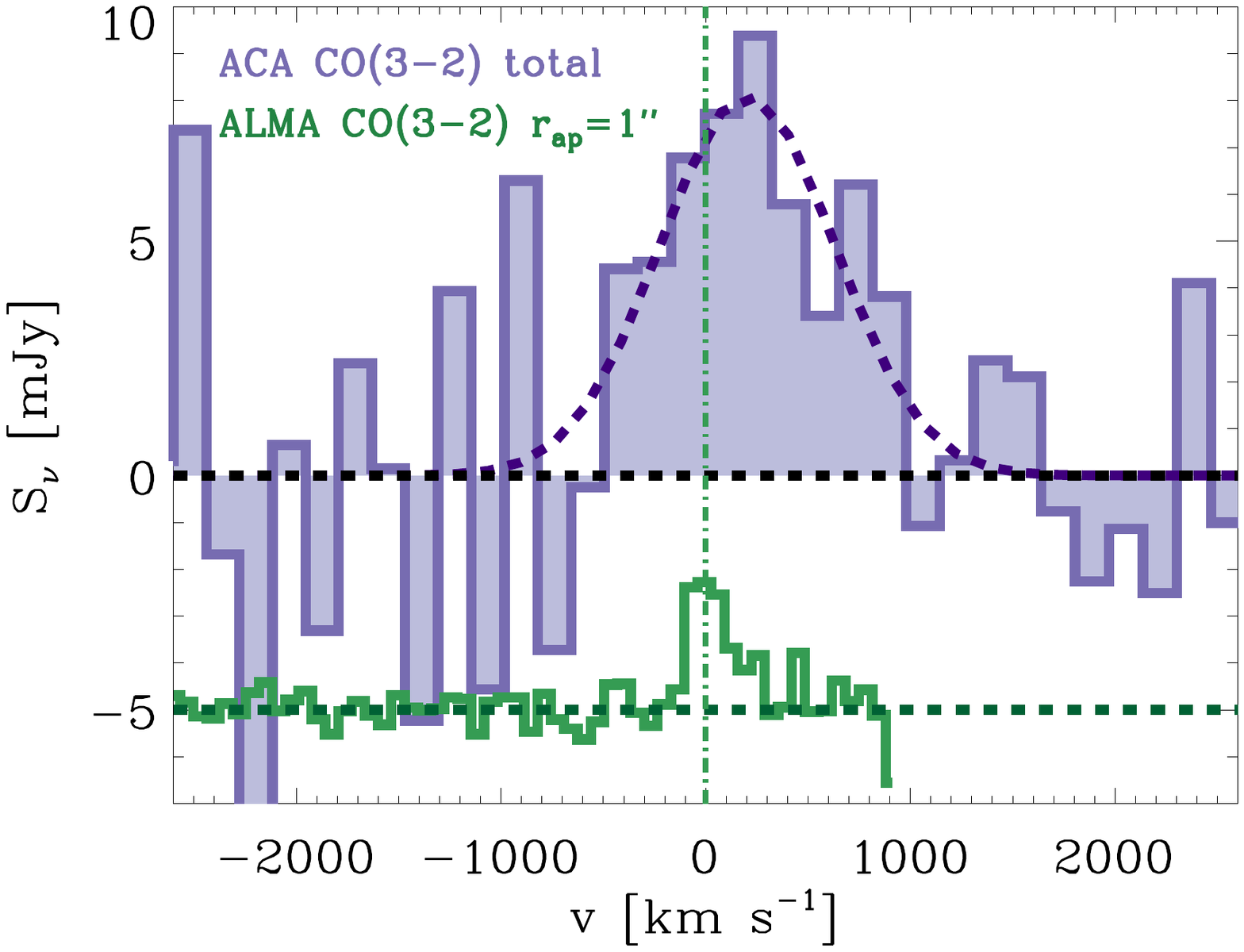}
        \caption{ACA CO(3-2) spectrum extracted from a polygonal region tracing the 2$\sigma$ contours in Fig.~\ref{fig:co32_aca_maps} which maximises the S/N (=5). The spectral profile is, however, almost identical to the ACA spectra extracted using circular or elliptical apertures with $r_{aperture}\gtrsim15''$ (Fig.~\ref{fig:co32_spectra_appendix}, bottom panel).   
        The spectral rms is 2.7~mJy per $\Delta v_{channel}=160$~km~s$^{-1}$. The best-fit Gaussian parameters are as follows: $v_{cen}=200\pm100$~\kms, FWHM= $1000\pm220$~\kms, and $S_{peak}=8.1\pm1.6$~mJy. At the bottom, shifted vertically to facilitate a visual comparison, we show the ALMA CO(3-2) spectrum extracted from a $1\arcsec$-radius circular aperture. The green dot-dashed line indicates the ALMA CO(3-2) line peak ($\nu_{obs}=107.4$~GHz, $z_{\rm CO}=2.2197$).
        }\label{fig:co32_spectra}
\end{figure}

\section{Results}

Figure~\ref{fig:co32_aca_maps} shows the ACA CO(3-2) map obtained by integrating the line emission across its full spectral extent of $-400<v[\rm km~s^{-1}]<1000$. 
The source was resolved at the ACA resolution of $9.3\arcsec\simeq80$~kpc, spreading across tens of arcsec in projected size. It shows a central peak, slightly ($\sim3\arcsec$) offset from the AGN position in the south-east direction, and additional secondary peaks north, north-east, and south-east of the nucleus. To our knowledge, this is the most extended molecular CGM reservoir that has ever been mapped.

The ACA CO(3-2) spectrum is shown in Figure~\ref{fig:co32_spectra}. 
A Gaussian fit returns a total flux of $8.7\pm2.6$~Jy~\kms, which is a factor of $14\pm5$ higher than estimated by \cite{Circosta+21} for the ISM component.
The additional CO(3-2) emission detected by ACA spans a much broader range in line-of-sight velocities than the inner ISM. Although our spectral analysis is limited by the low signal-to-noise (S/N) of the data,
by comparing ACA and ALMA spectra extracted from different apertures (see Figure~\ref{fig:co32_spectra} and additional spectra in Fig.~\ref{fig:co32_spectra_appendix}), we infer that: (i) the CO(3-2) emission within $r_{\rm ap}\lesssim 5\arcsec$ is dominated by a narrow (FWHM$\sim200$~\kms) line centred at $\nu_{\rm obs}=107.4$~GHz (corresponding to the $z_{\rm CO}$ listed in Table~\ref{table:properties}). Its peak flux density is maximised in the spectrum extracted from $r_{\rm ap}=2.5\arcsec$ ($S^{\rm peak}_{\nu}\sim4$~mJy), and it remains constant when using bigger extraction areas. (ii) Larger apertures collect significant additional flux due to a second line centred at $v\sim450$~\kms, a feature that is marginally detected by ALMA even in smaller apertures, but it only becomes dominant at $r_{\rm ap}>5\arcsec$. This component is responsible for the red shift of the main CO(3-2) peak in the low resolution ACA spectra. 
(iii) For $r_{\rm ap}>15\arcsec$, which are only reliably probed by ACA, we observed -- at a spectral resolution of $\Delta v_{channel}=160$~km~s$^{-1}$ -- a very broad line profile extending from $v\sim-400$~\kms~to $v\sim1000$~\kms with respect to a central frequency of $\nu_{\rm obs}=107.4$~GHz, with an additional fainter component at $v\sim1500$~\kms (see Fig.~\ref{fig:co32_spectra_appendix}, bottom panel).

\begin{figure}[tbp]
        \centering
        \includegraphics[clip=true,trim=2cm 3cm 3.5cm 3.2cm,scale=.44,angle=270]{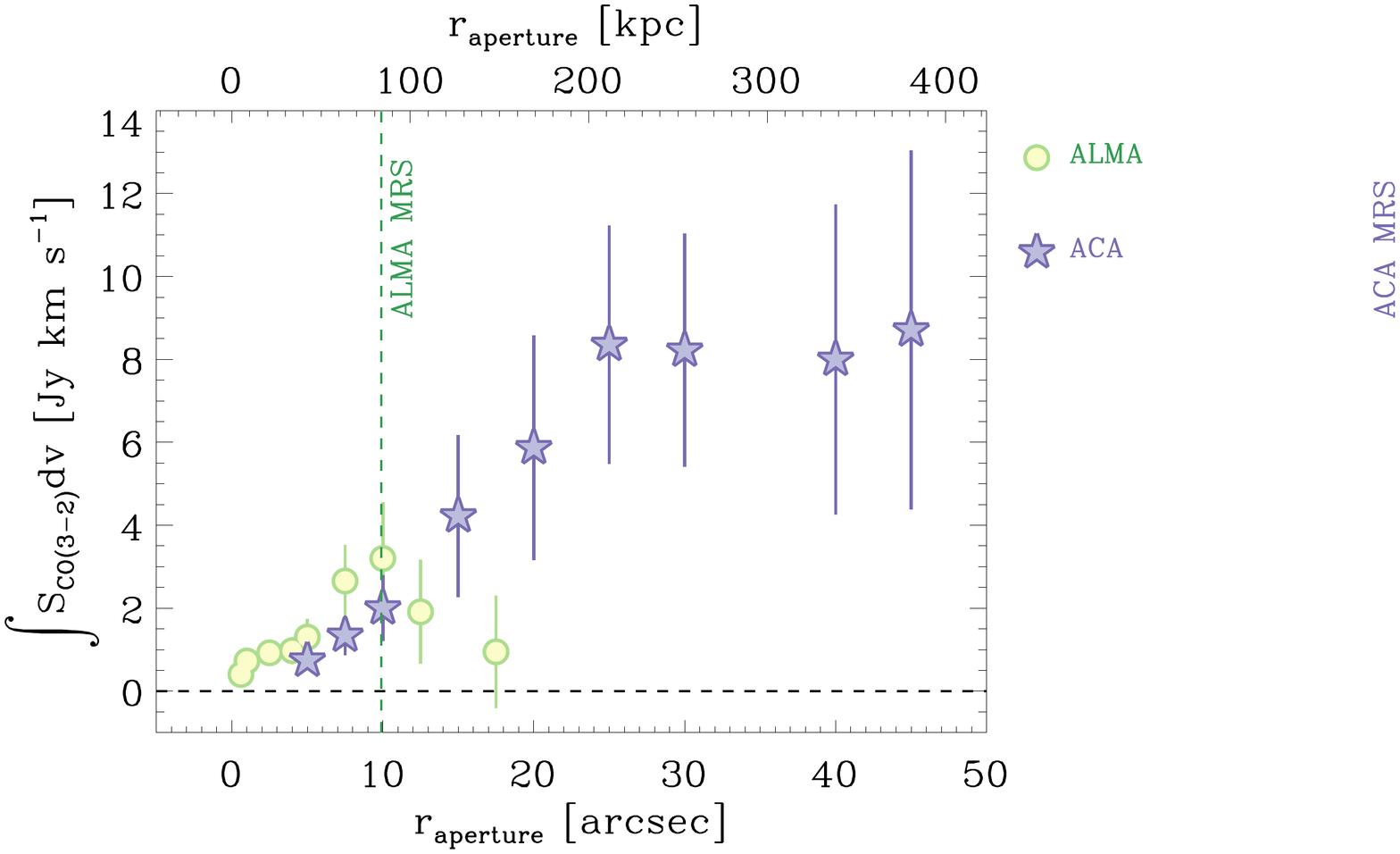}
        \caption{Curve of growth of the CO(3-2) line emission, showing the integrated flux as a function of the radius of the aperture used for spectra extraction. The fluxes were computed through a Gaussian fitting (details in Sect.~\ref{sec:appendix_co32_spectral_fit}). Vertical lines correspond to the MRS of the ALMA and ACA data. Error bars include nominal calibration uncertainties (5\% for both ALMA and ACA Band 3). Our ALMA data are not reliable for measuring fluxes beyond $r>10\arcsec$, where aperture dilution effects become severe due to the poor sensitivity to extended and redshifted components.
        }\label{fig:co32_flux_apertures}
\end{figure}

\begin{figure}[tbp]
        \centering
        \includegraphics[clip=true,trim=0cm 5.4cm 0cm 5.4cm,scale=.32,angle=0]{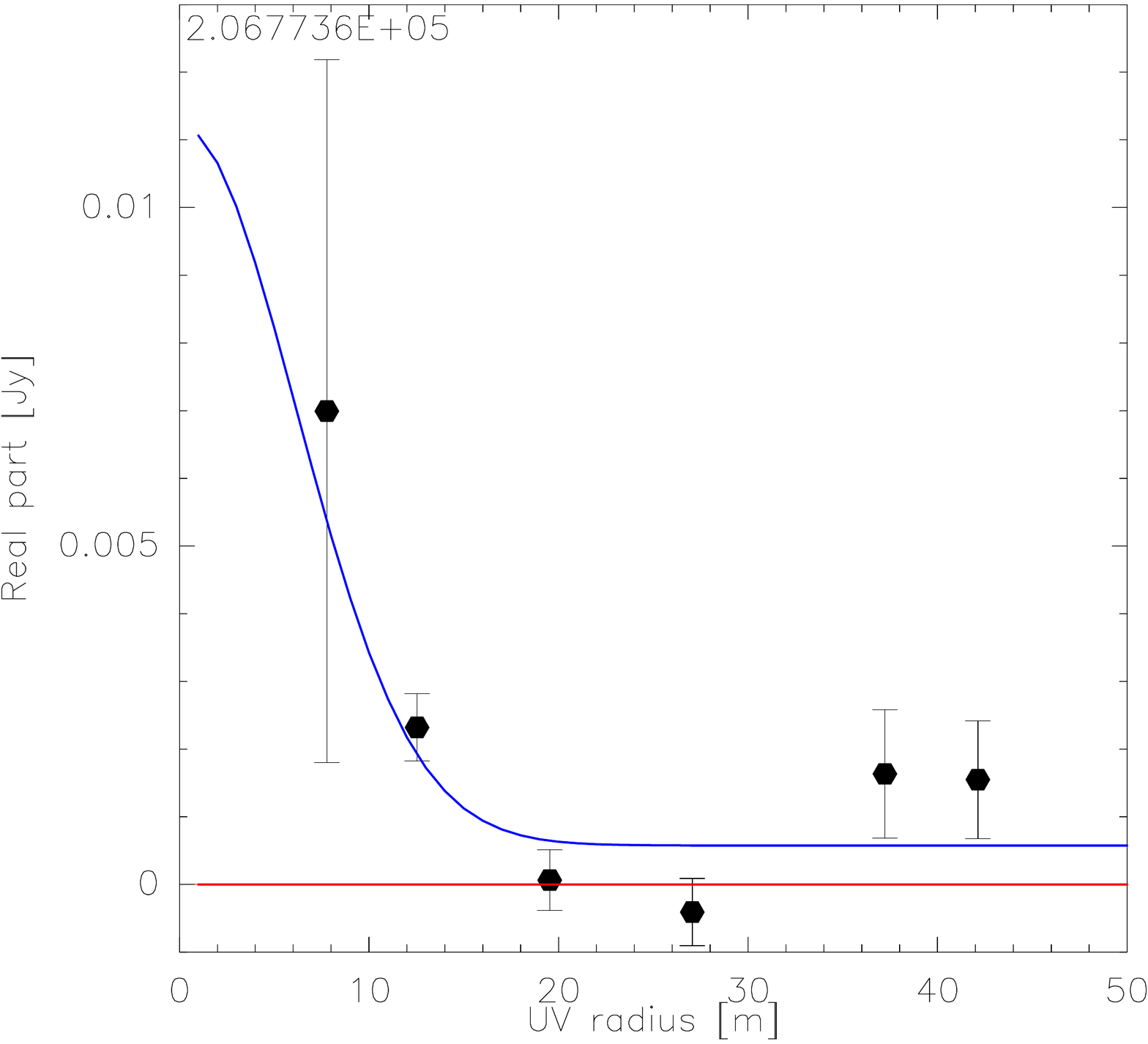}
        \caption{{\it uv} plot of the ACA CO(3-2) data integrated between $-400<v[\rm km~s^{-1}]<1000$ and binned in uv radii of 8~m. The blue curve is the best-fit with two components: a point source with a flux = $0.8\pm 0.5$~Jy~\kms, consistent with the fiducial ISM value, and a circular Gaussian with a flux = $15\pm11$~Jy~\kms and a FWHM = $35\arcsec\pm10\arcsec$.
        }\label{fig:co32_aca_uvplots}
\end{figure}

The curve of growth in Figure~\ref{fig:co32_flux_apertures} shows a steady increase in CO(3-2) flux up to $r_{\rm ap}\simeq 25\arcsec$, and it plateaus for larger apertures. From this, we infer that the CO halo extends up to $r\sim200$~kpc. 
This size estimate is consistent with the analysis of the {\it uv} visibilities presented in Fig.~\ref{fig:co32_aca_uvplots}, where the data are modelled using a circular Gaussian function with $\rm FWHM=300\pm80$~kpc, in addition to a point source (the unresolved ISM) which contributes very little to the total flux.

\section{Discussion and conclusions}\label{sec:discussion}

\begin{table*}[tbp]
   \centering 
        \caption{Summary of CO(3-2) measurements }
        \label{table:results}
        \begin{tabular}{lccl}
                \hline
                \hline 
                Region  &  $\int S_{CO(3-2)}dv$   & $\rm L^{\prime}_{CO(3-2)}$ & Notes \\
                                                &    [Jy~\kms]                                          & [10$^{10}$~K~km~s$^{-1}$~pc$^2$] & \\
                \hline 
                $r\lesssim8$~kpc &  $0.63\pm0.09$  & $1.74\pm0.24$ &  fiducial ISM \citep{Circosta+21} \\ 
                $r\lesssim200$~kpc  &  $8.7\pm2.6$ &  $24\pm7$  & fiducial CGM (Figure~\ref{fig:co32_spectra})  \\
                \hline 
                FWHM=$300\pm80$~kpc & $16\pm11$  &  $40\pm30$  &  {\it uv} fit (Figure~\ref{fig:co32_aca_uvplots})  \\  
                \hline
        \end{tabular}
        
        \begin{flushleft}
                {\it Notes:} 
            Errors include a 5\% calibration uncertainty.
        \end{flushleft}
\end{table*}

Table~\ref{table:results} summarises the CO(3-2) measurements available for cid\_346 and its halo: our fiducial ISM value is the one measured by \cite{Circosta+21} at $r\lesssim8$~kpc, while the fiducial CGM CO(3-2) flux (which includes the ISM contribution) is the one obtained through the Gaussian fit shown in Fig.~\ref{fig:co32_spectra}. In the same table, we also report the zeroth-baseline flux extrapolated through the {\it uv} visibility fit in Fig.~\ref{fig:co32_aca_uvplots}, which is consistent, within its (large) error, with the CGM flux measured from the spectrum. 

Any estimate of molecular gas mass based on a single CO transition is highly uncertain. The uncertainty is even higher in the absence of any (theoretical or observational) constraint on the physical conditions of the gas, which is the case for this newly discovered molecular CGM reservoir around cid\_346.  An extremely strict lower boundary on $M_{mol}^{CGM}$, obtained by simply counting the CO molecules in the $J=3$ level in an optically thin regime, is $M_{mol}^{CGM}\gg 2\times 10^9$~M$_{\odot}$ (see Sect~\ref{sec:mmol_limits}). A more reasonable (but still quite strict) lower limit is $M_{mol}^{CGM}> 10^{10}$~M$_{\odot}$, inferred assuming local thermodynamic equilibrium (LTE). There is however room for up to $1.7\times 10^{12}$~M$_{\odot}$ of molecular gas in the CGM of cid\_346 (upper limit assuming optically thick CO with $\alpha_{\rm CO}=3.6$~$\rm M_{\odot}~(K~km~s^{-1}~pc^2)^{-1}$ and $L^{\prime}_{CO(3-2)}/L^{\prime}_{CO(1-0)}=0.5$, see Sect~\ref{sec:mmol_limits}). As a comparison, it has been estimated that the CGM of the Milky Way contains a few 10$^9~M_{\odot}$ of gas within 75~kpc \citep{Zheng+19}, although a constraint on the molecular phase, as well as on the warm-hot medium, is still missing. 

The origin of such a massive molecular CGM reservoir around cid\_346 is unclear. Current data (see Sect.~\ref{sec:field_overdensity}) disfavour an over-density of galaxies, contrary to previous (albeit much less extended) detections of molecular halos, all in high-z protoclusters \citep{Ginolfi+17, Emonts+16}. Therefore, in the absence of better quality data, we set aside the hypothesis that this cold CGM reservoir coincides with a dense environment and thus that it may be the result of gravitational interactions and ram-pressure stripping from satellites.

The moment maps (see Section~\ref{sec:moment_maps}) show complex kinematics, which may be a signature of the direct imprint by AGN-driven outflows, but also of accreting streams. Interestingly, cid\_346 is one of the few Type 1 SUPER targets for which high velocity [OIII] $\lambda5007\AA{}$ emission, which is a signature of ionised outflows, has been detected as far as $\sim3$~kpc south-east of the AGN, using SINFONI-AO data \citep{Kakkad+20}. This is also the same direction of the offset of the global ACA CO(3-2) peak with respect to the AGN position (Fig.~\ref{fig:co32_aca_maps}), although we should not over-interpret this result, since the map in Fig.~\ref{fig:co32_aca_maps} results from an integration of the CO(3-2) emission over a very broad velocity range, and by restricting the velocity range the peak would shift (see e.g. Fig.~\ref{fig:co32_aca_moment_maps}, left panel). The large line width of the CO emission stemming from the CGM (Fig.~\ref{fig:co32_spectra}) is suggestive of outflows, but it may also simply reflect gas assembly within the dark matter halo. 

Understanding the presence and survival of molecular gas out to $r\sim200$~kpc is a real puzzle. On the one hand, the fact that the CGM of massive high-z galaxies is shaped by outflows, and, specifically, AGN-driven outflows, is a prediction of models \citep{Nelson+19, Suresh+19}. On the other hand, whether such outflows can have a direct impact on -- or even be responsible for -- a molecular CGM reservoir has yet to be proven from a theoretical perspective. 
It has been shown that clumps of cold gas can survive the entrainment by a hot medium if they are large enough (e.g. \citealt{Armillotta+17, Gronke+Oh18}). This process has, however, never been probed on CGM scales in a cosmological box.
In current simulations, even though AGN-driven outflows can have a higher cool ($T\sim10^4~K$) mass content and thus contribute to seed cold gas on large scales, the outer CGM is still predominantly hot \citep{Nelson+20}. This could be a numerical resolution bias \citep{Hummels+19} since no current cosmological simulation can track the survival or formation of molecular gas on scales of 100s of kiloparsecs. Observational evidence that outflows may be linked to extended cold gas CGM reservoirs has been building up, especially in powerful AGN \citep{Cicone+15, Travascio+20, Izumi+21}, and cid\_346 is the most extraordinary of such cases. 
However, an in-depth study of such extended cold CGM structures, especially at $z\sim0$, will only be possible with a facility such as the Atacama Large Aperture Submillimeter Telescope (AtLAST, e.g. \citealt{Klaassen+20, Cicone+19}).

\begin{acknowledgements}
We thank the anonymous referee for the very valuable feedback. CC thanks Padelis P. Papadopoulos for helping us derive sensible limits on the molecular gas masses, and Rolf G{\"u}sten for early access to the PI230 instrument. We warmly thank the APEX and ALMA staff for their technical support and their excellent service to the community, especially during the current pandemic. This paper makes use of the following ACA and ALMA data: ADS\textbackslash JAO.ALMA\#2019.2.00118.S, ADS\textbackslash JAO.ALMA\#2016.1.00798.S. ALMA is a partnership of ESO (representing its member states), NSF (USA) and NINS (Japan), together with NRC (Canada), MOST and ASIAA (Taiwan), and KASI (Republic of Korea), in cooperation with the Republic of Chile. The Joint ALMA Observatory is operated by ESO, AUI/NRAO and NAOJ. This publication is based on data acquired with the Atacama Pathfinder Experiment (APEX). APEX is a collaboration between the Max-Planck-Institut fur Radioastronomie, the European Southern Observatory, and the Onsala Space Observatory. Based on observations collected at the European Organisation for Astronomical Research in the Southern Hemisphere under ESO programmes 098.A-0774(A) and 0101.B-0758(A). 
GV acknowledges financial support from Premiale 2015 MITiC (PI: B. Garilli).
This project has received funding from the European Union’s Horizon 2020 research and innovation programme under grant agreement No 951815.
\end{acknowledgements}

\bibliographystyle{aa}
\bibliography{super}

\begin{appendix}
\section{Supplementary material}\label{sec:appendix}

\subsection{ACA non-detection of the continuum}\label{sec:continuum}

\begin{figure}[h]
        \centering
        \includegraphics[clip=true,trim=2.cm 7.7cm 2.cm 1.4cm,width=.9\columnwidth,angle=0]{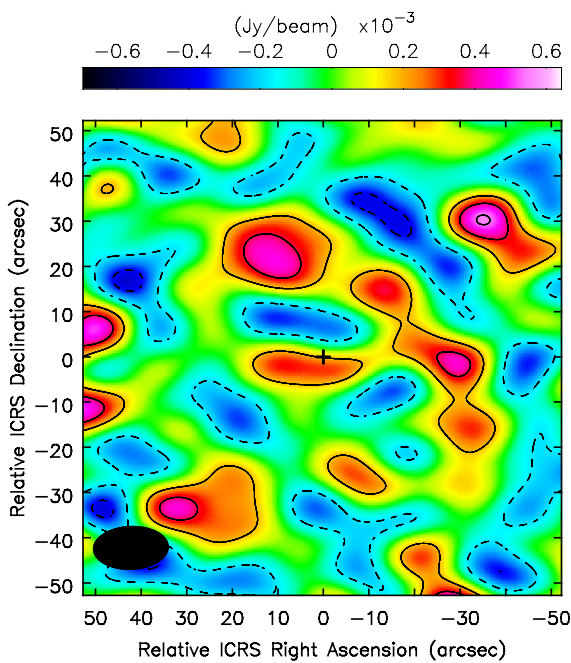}
        \caption{ACA map showing the non-detection of the 3mm continuum in cid\_346, obtained by integrating the emission over the (observed) frequency range $\nu_{obs}\in(104.8, 106.75)$~GHz. Negative and positive contours are plotted in steps of $1\sigma$ starting from $\pm1\sigma$, with a $1\sigma=0.18$~mJy~beam$^{-1}$. }
        \label{fig:ACA_cont}
\end{figure}

The 3mm continuum flux density measured in cid\_346 by \cite{Circosta+21} using ALMA is $0.15\pm0.04$~mJy. Such continuum emission is not detected by ACA. We verified this by using a line-free spectral window (spw17) adjacent to the one containing the CO(3-2) line, but at a lower spectral resolution of $\Delta v_{\rm channel}=22$~\kms, which covers observed frequencies between 106.75 and 104.80 GHz. In Figure~\ref{fig:ACA_cont} we present a collapsed map of the full spw17 range, centred at $\nu_{obs}=105.796$~GHz, which does not show any significant detection. Using this map, we can place an ACA $3\sigma$ upper limit on the continuum level of $\rm S_{106~GHz}<0.5~mJy~beam^{-1}$, which is consistent with the detection by \cite{Circosta+21}.

\subsection{The APEX [CI](2-1) data}\label{sec:apex_spec}
\begin{figure}[h]
        \centering
        \includegraphics[clip=true,trim=2cm 0cm 2.cm 0cm, scale=.3,angle=270]{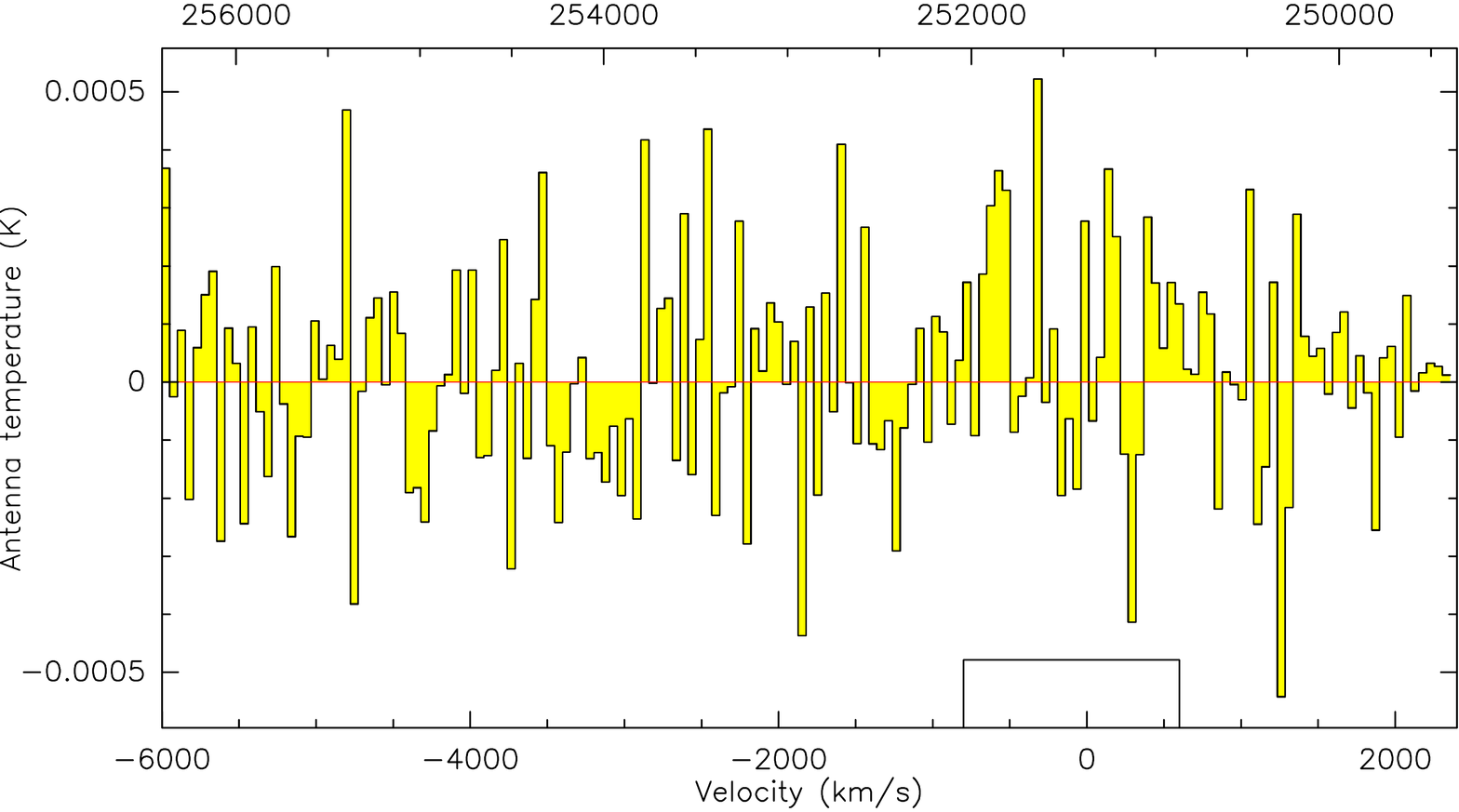}
        \caption{APEX [CI](2-1) spectrum, with no signs of either the [CI] or the CO emission lines. The window used to mask the line during the baseline fitting is shown. The y-axis units are antenna temperature corrected for atmospheric losses ($T_A^{\prime}$). The final rms calculated across the full spectral range ($v\in(-6000, 2400)$~\kms) is 0.175~mK per $\Delta v=50$~\kms~channel.}
        \label{fig:APEX_CI_spec}
\end{figure}
Figure~\ref{fig:APEX_CI_spec} shows the final APEX PI230 spectrum, obtained by combining all the good quality scans selected from the 2017 and 2018 observing runs, for a total of 16.9 hours of single-polarisation data. 
The tuning frequency is the redshifted frequency of the [CI](2-1) line ($\rm \nu_{obs}=251.372$~GHz, set to $v=0$~\kms). The CO(7-6) line would be expected at $\rm \nu_{obs}=250.5363$~GHz, thus $\Delta v\sim 1000$~\kms~redward of the [CI](2-1) line. Neither of the two lines is detected in the APEX data, down to a sensitivity of $7.0\pm1.2$~mJy per $\Delta v=50$~\kms~channel.

By assuming that the [CI](2-1) and CO(3-2) lines share the same line profile, we can assume ${\rm FWHM}\sim1000$~\kms~(see Fig.~\ref{fig:co32_spectra}) to place a 3$\sigma$ upper limit on the [CI](2-1) flux:
\begin{equation}
        \rm \int S_{\nu}^{[CI]}dv < 3 \sigma_{rms}\sqrt{\Delta v_{channel}\cdot {\rm FWHM}_{\rm [CI]}} \sim  4.7~Jy~km~s^{-1},
\end{equation}
which corresponds to $L^{\prime}_{\rm [CI](2-1)}<2.4\times10^{10}$~K~\kms~pc$^2$.

\subsection{CO(3-2) line spectral fits}\label{sec:appendix_co32_spectral_fit}

\begin{figure}[tb]
        \centering
        \includegraphics[clip=true,trim=3cm 2.cm 1.4cm 2cm,scale=.34,angle=270]{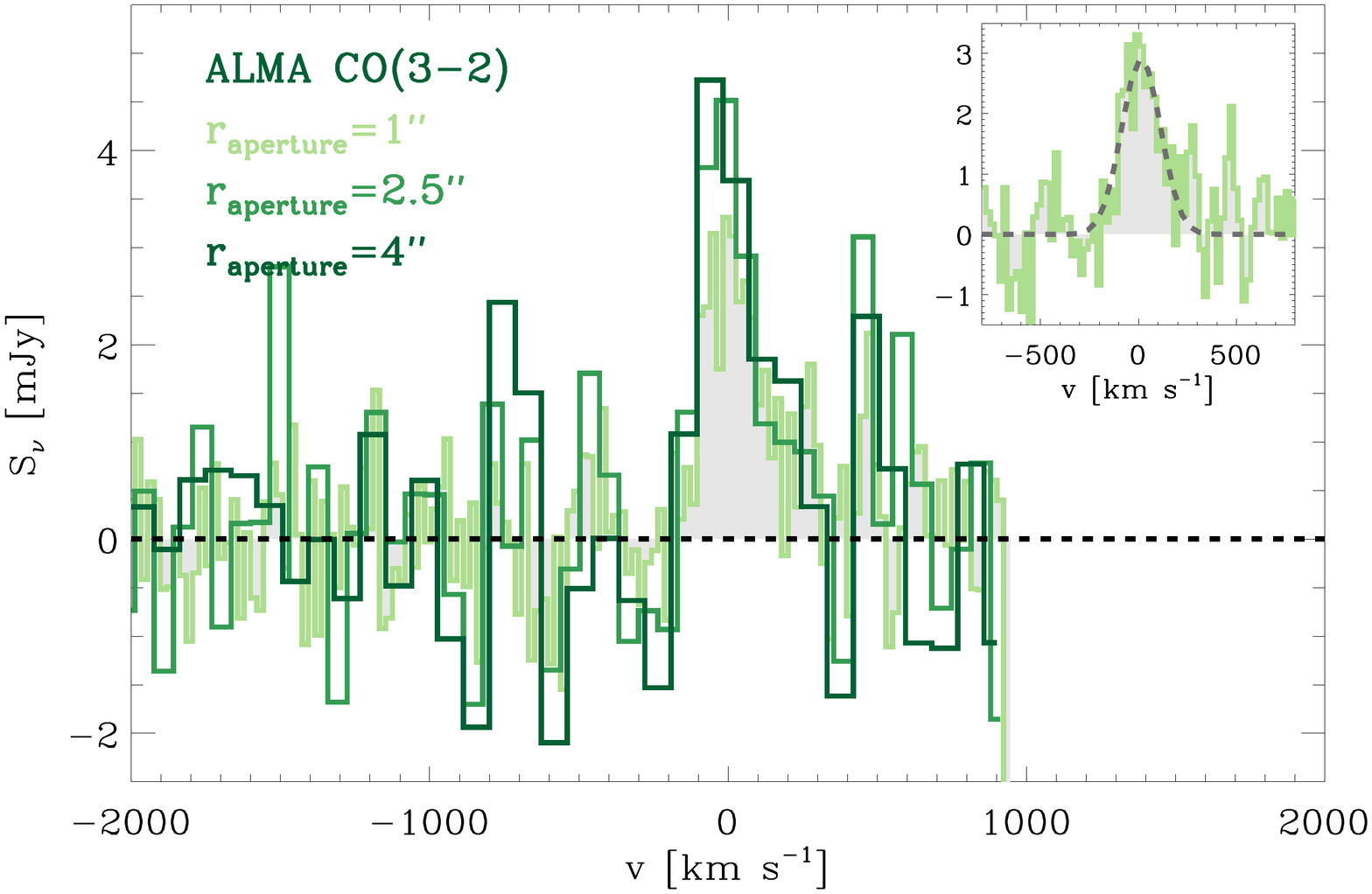}
	
        \includegraphics[clip=true,trim=3cm 2.cm 1.4cm 2cm,scale=.34,angle=270]{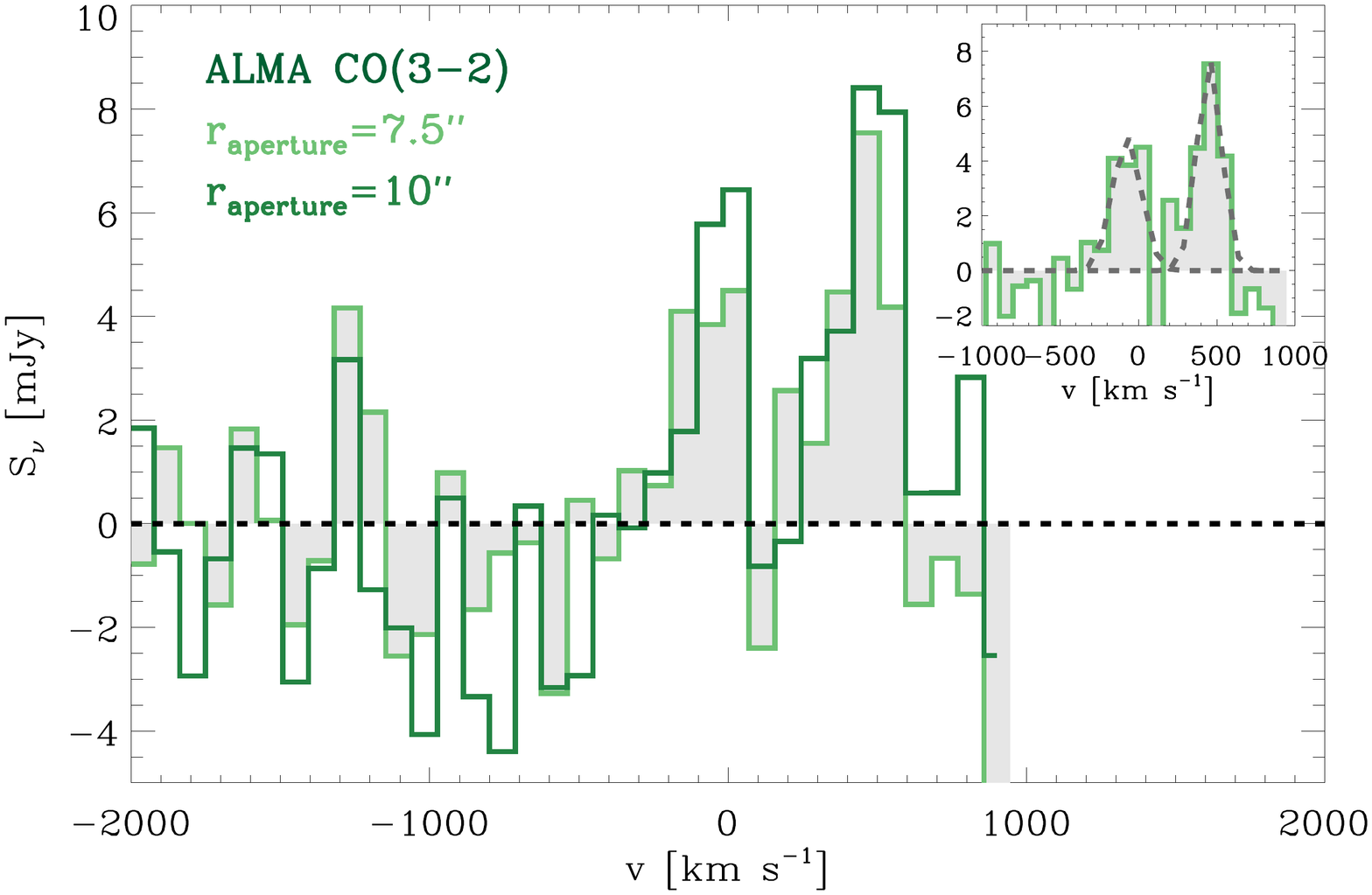}
	
        \includegraphics[clip=true,trim=3cm 2cm 1.4cm 2cm,scale=.34,angle=270]{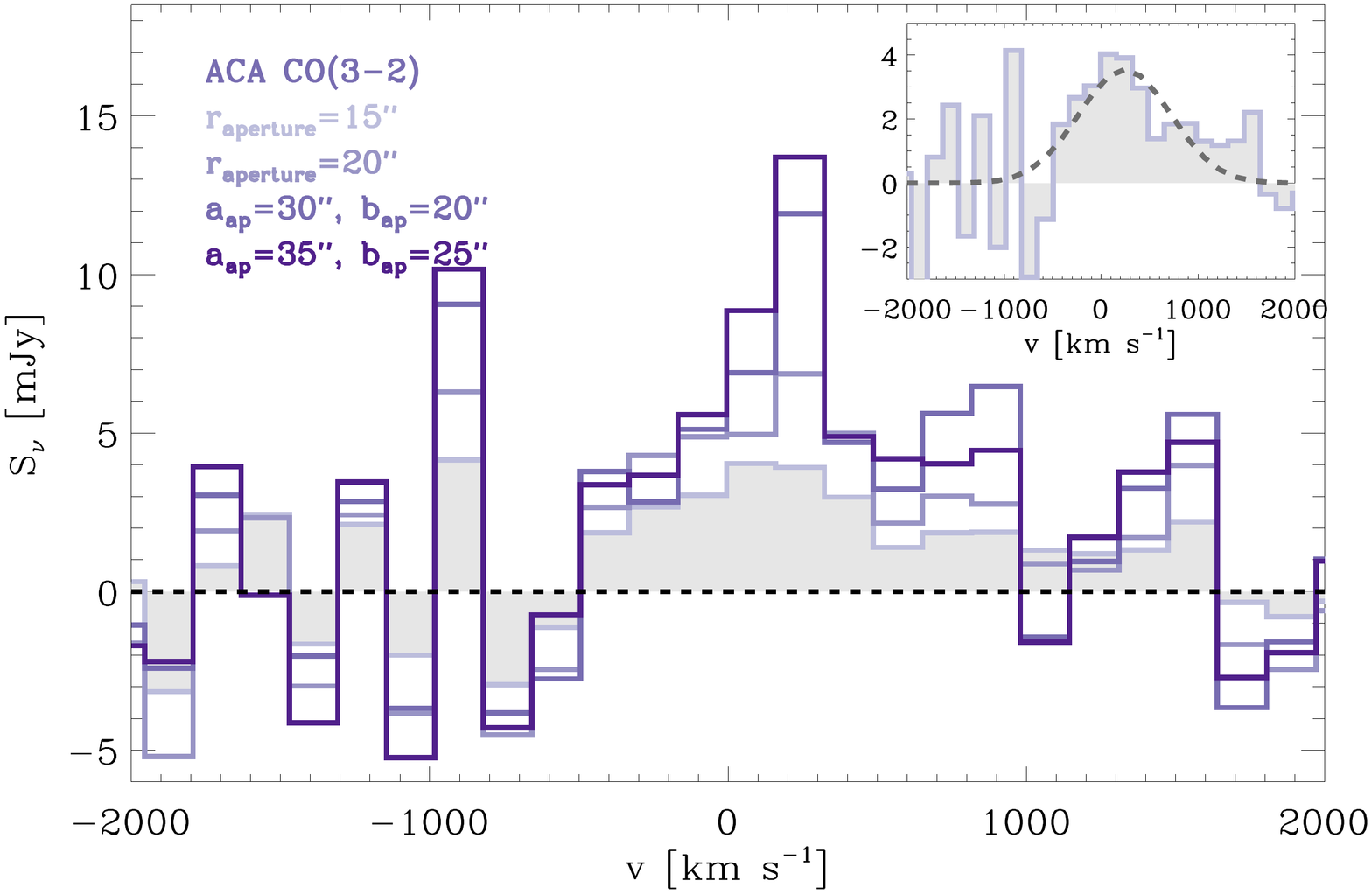}
        \caption{CO(3-2) spectra extracted using increasingly larger apertures from the primary-beam corrected ALMA and ACA data cubes. The channel widths are 21, 64, 86~km~s$^{-1}$ for the ALMA $r=1\arcsec$, $2.5\arcsec$, $4\arcsec$ spectra ({\em top panel}), 86~\kms~for the ALMA $r=7.5\arcsec$ and $10\arcsec$ spectra ({\em middle panel}), and $160$~km~s$^{-1}$ for the ACA spectra ({\em bottom panel}). The aperture radii (or, in the case of elliptical apertures, the major and minor semi-axes) are reported on the plots. The top-right insets in each panel show the Gaussian spectral fits to the ALMA $r_{\rm ap}=1\arcsec$ ({\em top}),  ALMA $r_{\rm ap}=7.5\arcsec$ ({\em middle}), and ACA $r_{\rm ap}=15\arcsec$ ({\em bottom}) spectra.
        }\label{fig:co32_spectra_appendix}
\end{figure}

In this section we describe, in more detail, the Gaussian fitting procedure used to measure the CO(3-2) fluxes for the curve of growth reported in Fig.~\ref{fig:co32_flux_apertures}. A few representative spectra used to compute Fig.~\ref{fig:co32_flux_apertures} are reported in Fig.~\ref{fig:co32_spectra_appendix}. 
The CO(3-2) spectral line profile significantly varies with the aperture used for spectra extraction, and of course it also depends on the S/N of the data and spectral binning applied. The spectra used in this analysis can be divided into three categories, shown in different panels of Fig.~\ref{fig:co32_spectra_appendix}. 

For small apertures probed by the ALMA data ($r_{aperture}=0.6\arcsec-5\arcsec$), the CO(3-2) spectrum exhibits a single component, which can be modelled using a Gaussian function. The top panel of Fig.~\ref{fig:co32_spectra_appendix} shows a few examples and reports, in the top-right inset, the best fit to the $r_{aperture}=1''$ spectrum.

The ALMA spectra extracted using $r_{aperture}=7.5\arcsec$ and $10\arcsec$ show a clear additional redshifted component at $v\sim 450-480$~\kms, brighter than the one at $v=0$~\kms,  which is very close to the edge of the ALMA spectral window. These spectra are shown in the middle panel of Fig.~\ref{fig:co32_spectra_appendix}.

The ALMA spectra extracted from even larger apertures, with $r_{aperture}>10\arcsec$, are not reliable, even when using the tapered dataset (ALMA-t in Table~\ref{table:properties}). Indeed, at these scales, the snapshot $1\arcsec$-resolution ALMA data fail to recover the additional extended flux and the increasingly larger contribution from the redshifted component (due to their limited bandwidth on the red side of the line), and the resulting flux measurement is affected by severe aperture dilution effects. These technical issues are instructively shown by the two ALMA data points at $r_{aperture}>10\arcsec$ in Fig.~\ref{fig:co32_flux_apertures}. These fluxes were obtained by fitting the spectra extracted from the ALMA-t cube, using elliptical apertures with semi-major axes of $a=15\arcsec$, $b=10\arcsec$ for the data point plotted at $r_{aperture}=12.5\arcsec$, and 
$a=20\arcsec$, $b=15\arcsec$, for the data point plotted at 
$r_{aperture}=17.5\arcsec$.
The resulting flux values fall below the $r_{aperture}=10\arcsec$ value, hence demonstrating the inadequacy of this dataset for probing extended structures.

Apertures with $r>10\arcsec$ are only reliably probed by the ACA data. The ACA spectra were rebinned to $\Delta v_{channel}=160$~\kms~in order to maximise the S/N. All spectra were fitted using a single broad Gaussian component, as shown in the bottom panel of Fig.~\ref{fig:co32_spectra_appendix} (see also Fig.~\ref{fig:co32_spectra}).
The central velocity of the best-fit Gaussian function varies within the range $v\sim200-300$~\kms~for different apertures, hence robustly displaying a red shift compared to the CO(3-2) emission arising from the inner ISM of cid\_346. This is also clearly shown by the comparison between ALMA and ACA spectra in Fig.~\ref{fig:co32_spectra}. 
Although the low S/N of the ACA spectra does not allow us to analyse them at higher spectral resolution, it is likely that the observed global redshift of the CO(3-2) line results from the spectral blending (when using such a large bin size of $\Delta v_{channel}=160$~\kms) of the two peaks that are already visible in the ALMA spectra extracted from $r_{aperture}=7.5\arcsec$ and $10\arcsec$ (middle panel of Fig.~\ref{fig:co32_spectra_appendix}).
The best-fit FWHM value is $\sim1000$~\kms~for all ACA spectral fits. None of these single-Gaussian fits include the additional component centred at $v\sim1500$~\kms, which is detected at too low of an S/N to be fit separately, but it is consistently present in all the ACA spectra, as shown in the bottom panel of Fig.~\ref{fig:co32_spectra_appendix}. We note that we cannot verify whether this component is also present on the smaller ISM scales probed by ALMA because our ALMA data lack the corresponding frequency coverage.

\subsection{Lower and upper limits on $M_{mol}^{CGM}$}\label{sec:mmol_limits}

 In the following, we use our fiducial estimate of the CGM CO(3-2) luminosity, $L^{\prime}_{\rm CO(3-2)}=(24\pm7)\times10^{10}$~K~km~s$^{-1}$~pc$^2$ (Table~\ref{table:results}), to place strict lower and upper limits on the CGM molecular gas mass. We treat all of the CO(3-2) emission equally, although the physical conditions in the ISM (inner kiloparsecs) can be different from those on larger scales. In the absence of additional constraints, any assumption would be speculative.

\subsubsection{Lower limit on $M_{mol}^{CGM}$}

A lower boundary for $M_{mol}^{CGM}$ can be estimated using the optically thin limit. Even in this case, there are several possible routes (see, e.g. \citealt{Papadopoulos+12b}), which are more or less conservative depending on the underlying assumptions. The most minimalist one is to estimate the column density of the molecular gas in the $J=3$ level, $N_3$, and from this infer the molecular gas mass assuming only the $J=3$ level contribution, that is
\begin{equation}\label{eq:N3_contri}
 M_{mol}^{CGM} \gg N_3~R_{\rm CO}~\mu~m_{H_2} \sim 2 \times 10^9 ~M_{\odot},
\end{equation}
where $R_{CO}$ is the $\rm [H_2/CO]$ abundance ratio and $\mu=1.36$ accounts for the mass in Helium. We have assumed a typical CO abundance $\rm [CO/H_2]=10^{-4}$ since lower values, for example due to lower metallicities and/or CO-dark H$_2$ gas, would only increase the mass. We stress that Eq.~\ref{eq:N3_contri} represents an extremely strict lower limit (hence the $\gg$ symbol) that is physically unattainable because it does not account for the (significant) contribution to the total column density $N_{tot}$ due to the CO molecules in the $J=0,1,2$ levels (or in the $J\geq4$ levels). 

To account for the contribution from the other CO energy levels, we can assume local thermodynamic equilibrium (LTE) and so describe the system as a Boltzmann distribution with a single temperature, $T_{kin}$. Under this condition, the total column density $N_{tot}$ can be derived using
\begin{equation}
        \frac{N_{3}}{N_{tot}} =\frac{g_3}{Z_{\rm LTE}}e^{-E_{3}/ (k_B T_{kin})},
\end{equation}
where $g_3 = 7$, $Z_{LTE}$ is the LTE partition function given by $Z_{\rm LTE}\simeq 2k_B T_{kin}/E_1$, and the $J=1$ and $J=3$ state energies are $E_1/k_B = 5.53$~K and $E_3/k_B = 33.19$~K. Hence, a more reasonable lower limit for the CGM molecular gas mass is as follows:
\begin{equation}\label{eq:N3_LTE}
        M_{mol}^{CGM} > N_{tot} R_{\rm CO}~\mu~m_{H_2} \sim  10^{10}~M_{\odot}.
\end{equation}
Here we have assumed $T_{kin}=50$~K, which corresponds to the lower boundary of the cold neutral medium temperature range.

\subsubsection{Upper limit on $M_{mol}^{CGM}$}

We may estimate an upper limit on $M_{mol}^{CGM}$ by assuming optically thick CO emission throughout the entire CGM reservoir. Several recent literature studies promote the use of a CO-to-$H_2$ conversion factor of $\alpha_{\rm CO(1-0)}=3.6$~M$_{\odot}~(K~km~s^{-1}~pc^2)^{-1}$ (including the Helium correction) for the molecular ISM of $z\sim2$ galaxies (see references in \citealt{Circosta+21}). Concerning the CO excitation, the literature reports a broad range of $r_{31}=L^{\prime}_{CO(3-2)}/L^{\prime}_{CO(1-0)}$ measurements, $r_{31}\sim0.5-1$, at both low and high redshifts. If we assume $r_{31}\sim0.5$, combined with the $\alpha_{\rm CO(1-0)}$ quoted above, we obtain a total molecular gas mass of $1.7\times10^{12}~M_{\odot}$, which may be considered an upper limit given the premise of optically thick CO emission and the low $r_{31}$ ratio. 

Alternatively, it is possible to convert the APEX [CI](2-1) non-detection into an upper limit on  $M_{mol}^{CGM}$, by assuming a Carbon excitation temperature $T_{ex}$ and a Carbon abundance $X_{CI}=[C]/[H_2]$. The most updated high-z compilations of [CI](2-1) and [CI](1-0) line measurements by \cite{Valentino+18} and \cite{Valentino+20} provide average values of $L^{\prime}_{\rm [CI](2-1)}/L^{\prime}_{\rm [CI](1-0)}= 0.47$ (corresponding to $T_{ex}=26$~K) and $X_{\rm CI}= 1.6\times 10^{-5}$ for the ISM of main sequence galaxies at $z>1$. If we adopt these values we obtain $M_{mol}< 9 \times 10^{11}~M_{\odot}$, which is slightly more stringent than the $\alpha_{\rm CO(1-0)}$-based upper $M_{mol}$ boundary. However, we note that the APEX single-pointing FoV of $27\arcsec$ (Table~\ref{table:obs}) is smaller than the full extent of the CO halo detected by ACA, and an aperture of $r_{ap}=15\arcsec$ contains half of our fiducial CGM CO(3-2) flux ($4.2\pm1.9$~Jy~\kms, see Fig~\ref{fig:co32_flux_apertures}), hence the two upper limits are consistent.

Finally, we note that some caution must be taken when comparing these mass estimates with the previous molecular CGM detections by \cite{Ginolfi+17} and \cite{Emonts+16}. Indeed, these studies used very high $\alpha_{\rm CO(1-0)}$ values of $\alpha_{\rm CO(1-0)}=10$ and $\alpha_{\rm CO(1-0)}=4$~M$_{\odot}~(K~km~s^{-1}~pc^2)^{-1}$,  respectively.

\subsection{Continuum sources in the field}\label{sec:field_overdensity}

\begin{figure}[tbp]
        \centering
        \includegraphics[clip=true,trim=0cm 0cm 0cm 0cm,width=.9\columnwidth,angle=0]{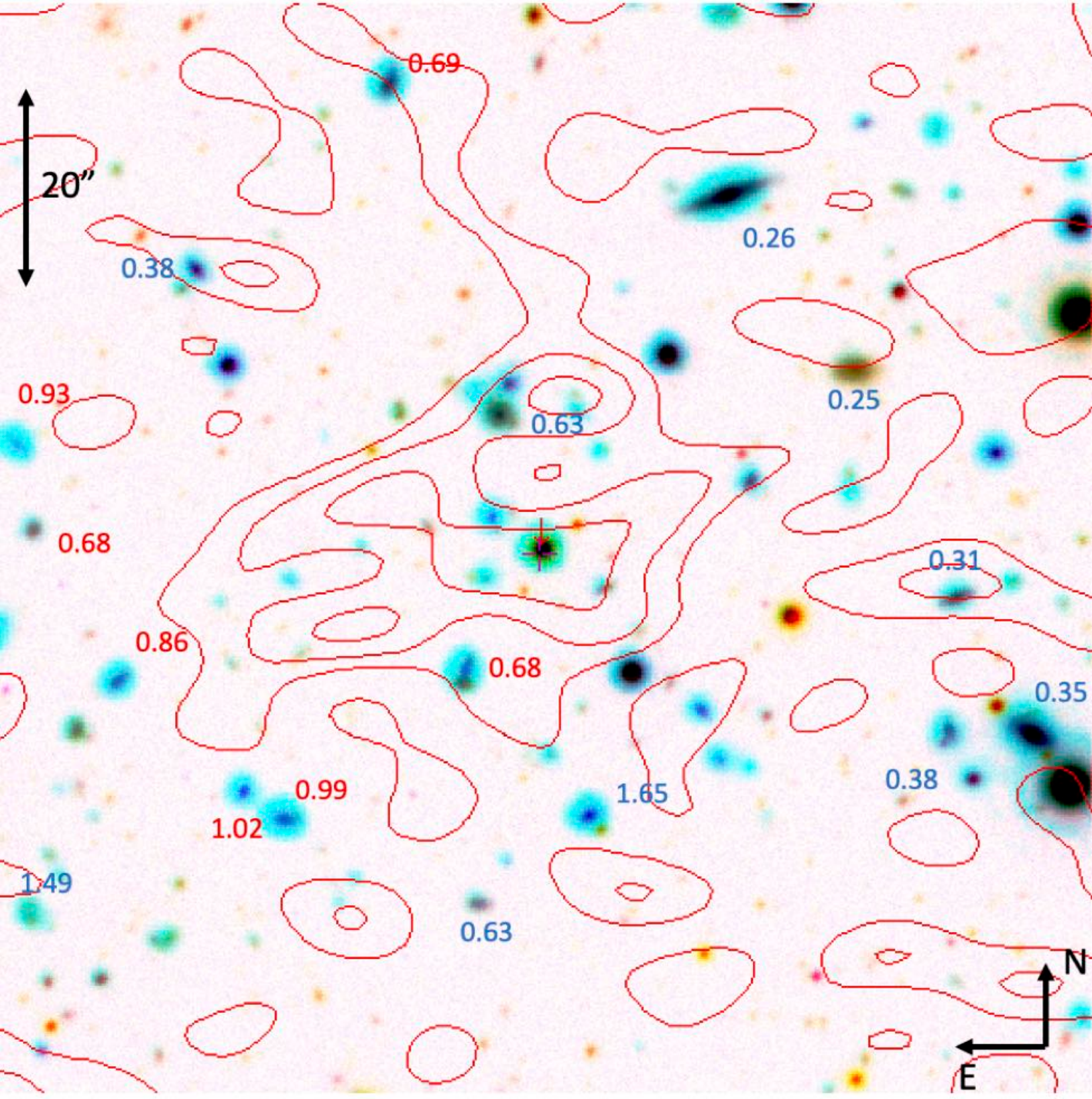}
        \caption{Optical and infrared image of the field around cid\_346, obtained by combining Subaru and Spitzer photometric data (see Table~\ref{table:RGB}). The images have been registered to the same astrometric solution, using the one with the best point spread function (z-band) as a reference. Known spectroscopic redshifts from the literature (blue) and photometric redshifts from \cite{Laigle+16} (red) are reported in the image. The red contours show the ACA CO(3-2) 1-3$\sigma$ level emission from Fig.~\ref{fig:co32_aca_maps}.
        }\label{fig:RGB}
\end{figure}

\begin{table*}
        \centering \small
        \caption{Description of the RGB data used in Fig.~\ref{fig:RGB}}
        \label{table:RGB}
        \begin{tabular}{llcccc}
                \hline
                \hline 
                Instrument & Band & $\lambda_{cen}$ & FWHM & PSF & 5$\sigma$-depth \\
                                &                                       &       [$\mu$m]                        & [$\mu$m]        & [$''$] & [mag]  \\
                \hline 
                Subaru/HSC$^{\dag}$  & g-band (Blue) & 0.4816 & 0.1386 & 0.63 & 26.84 \\
                Subaru/HSC$^{\dag}$  & z-band (Green) & 0.8912 &0.0773 & 0.45 & 24.79 \\
                Spitzer$^{\ddag}$ & IRAC ch2 (Red)  & 4.5049 & 1.0097 &1.4  & 25.13 \\
                \hline 
        \end{tabular}

        \begin{flushleft}
        {\it Notes:} 
        $^{\dag}$ \cite{Laigle+16};
        $^{\ddag}$ \cite{Mehta+18}.
\end{flushleft}
\end{table*}

 Figure~\ref{fig:RGB} displays the ACA CO(3-2) contours overlaid on the optical and infrared image of the field, obtained by combining photometric data from the {\it Subaru} and {\it Spitzer} telescopes. The central wavelength ($\lambda_{cen}$) and FWHM of the photometric filters, as well as the point spread function (PSF) and sensitivity of the images are listed in Table~\ref{table:RGB}. Figure~\ref{fig:RGB} shows that the ACA CO peaks do not coincide with any known optical or infrared source. The galaxies with a photometric or spectroscopic redshift measurement that lie close to the CO contours have redshifts that are not consistent with the range inferred from the CO data of $z\sim 2.215-2.230$. In addition, sensitive ALMA Band~7 continuum observations  \footnote{Project number 2018.1.00992.S, PI: C. Harrison} do not show any dust continuum emitters besides cid\_346 within a FoV of $16\arcsec$.
 
 Furthermore, there is no clear indication that cid\_346 is in a galaxy pair or in a merging state. The target appears as a point source both in the COSMOS HST-Advanced Camera for Surveys data (rest-frame UV, PSF=$0.12''$) and in the UltraVista near-infrared data (seeing-limited, FWHM $\simeq0.78''$) \citep{Laigle+16}. Even at longer wavelengths, cid\_346 does not show close companions. Its rest frame 260-$\mu$m continuum emission ($870\mu$m observed frame)  was imaged at $\sim2$~kpc resolution with a very high S/N ($\sim29$, \citealt{Lamperti+21}) and resolved into a Gaussian distribution with half-light radius $R_e=1.81$~kpc and a possible additional central point source (a compact starburst or the AGN) contributing $\sim13$\% to the flux.
 
 \subsection{CO(3-2) moment maps}\label{sec:moment_maps}

 \begin{figure*}[tb]
        \centering
        \includegraphics[clip=true,trim=1.5cm 7.5cm 2.4cm 1.2cm,scale=0.33,angle=0]{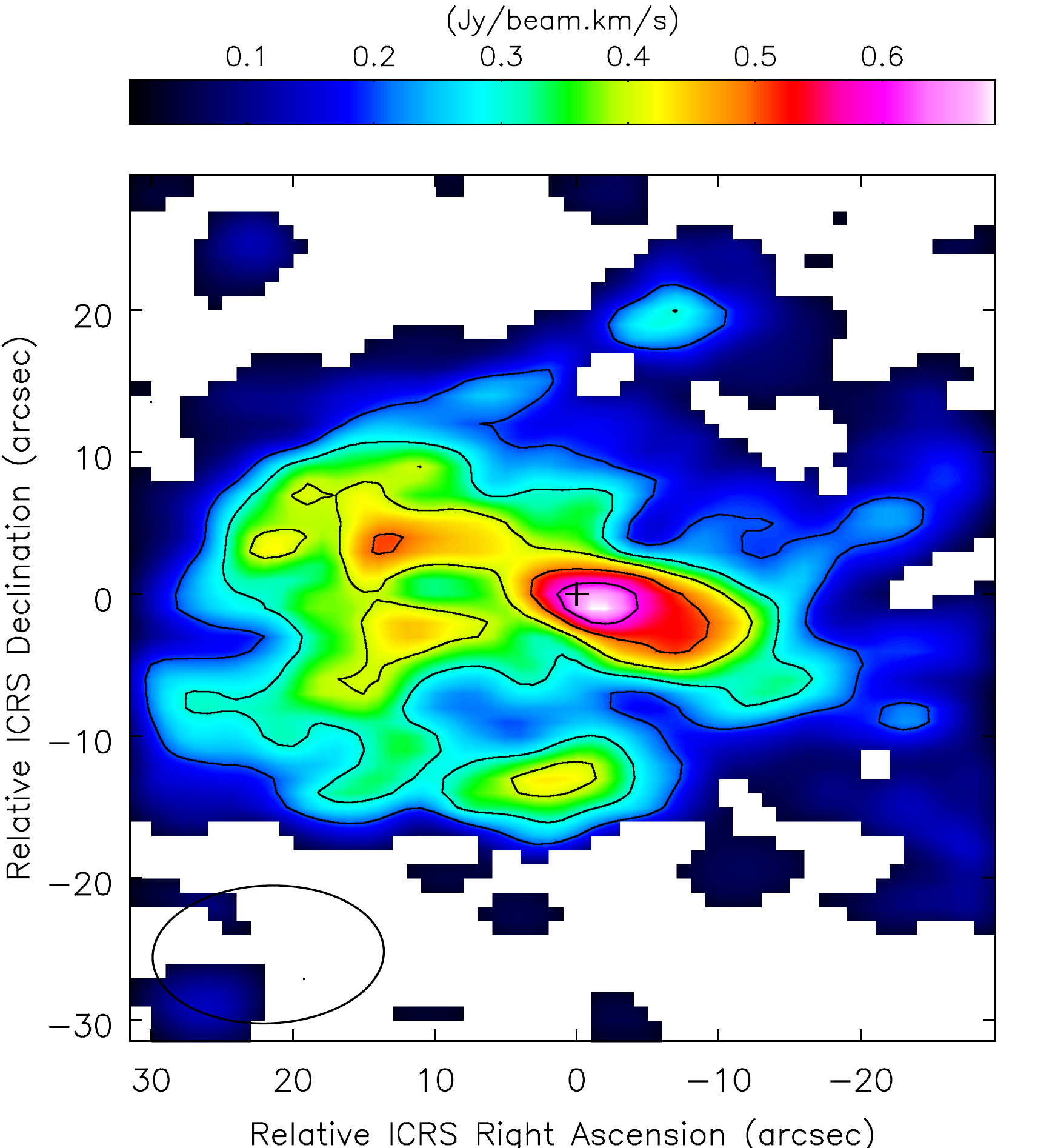}\quad
        \includegraphics[clip=true,trim=1.5cm 7.5cm 2.4cm 1.2cm,scale=0.33,angle=0]{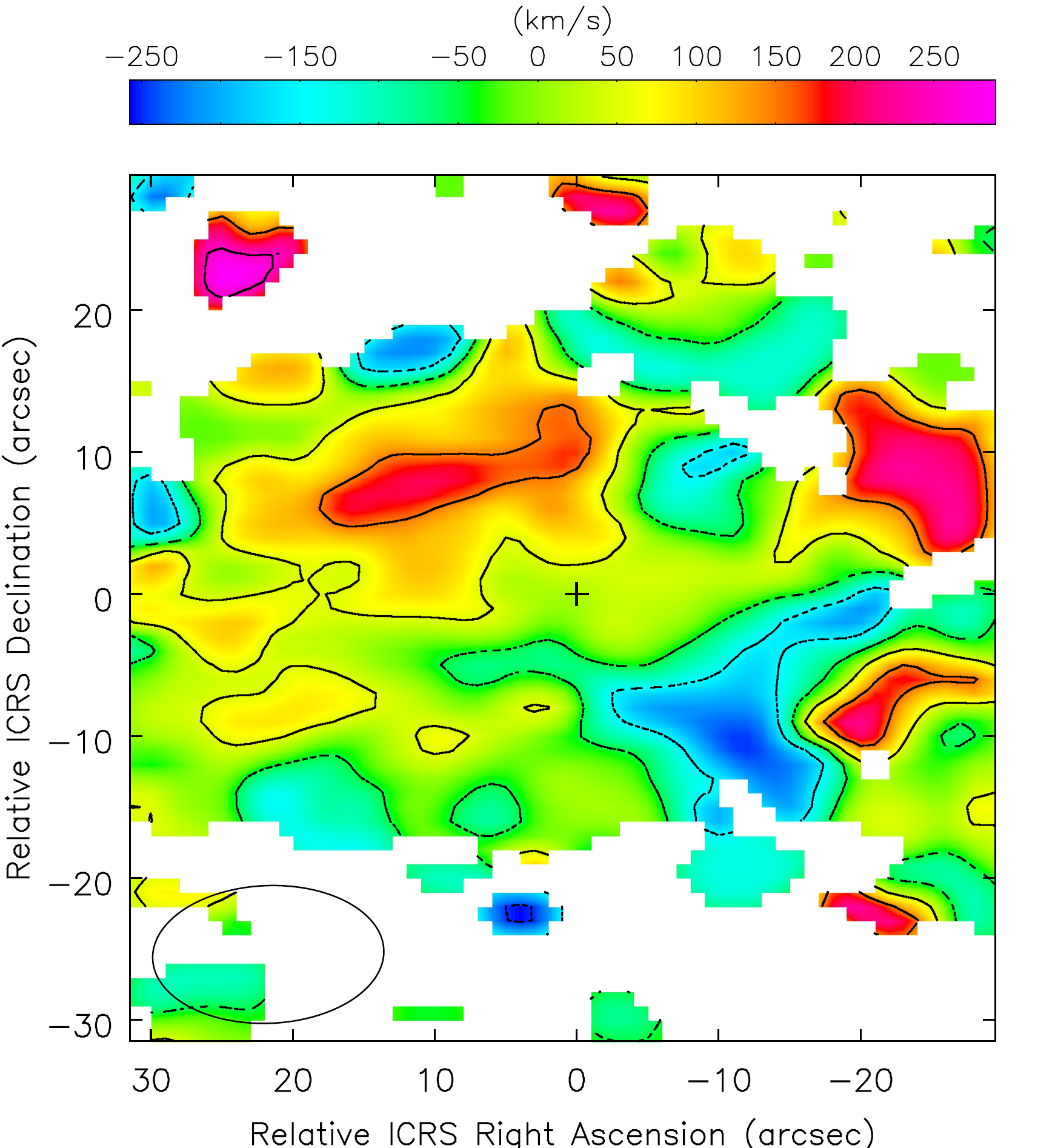}\quad
        \includegraphics[clip=true,trim=1.5cm 7.5cm 2.4cm 1.2cm,scale=0.33,angle=0]{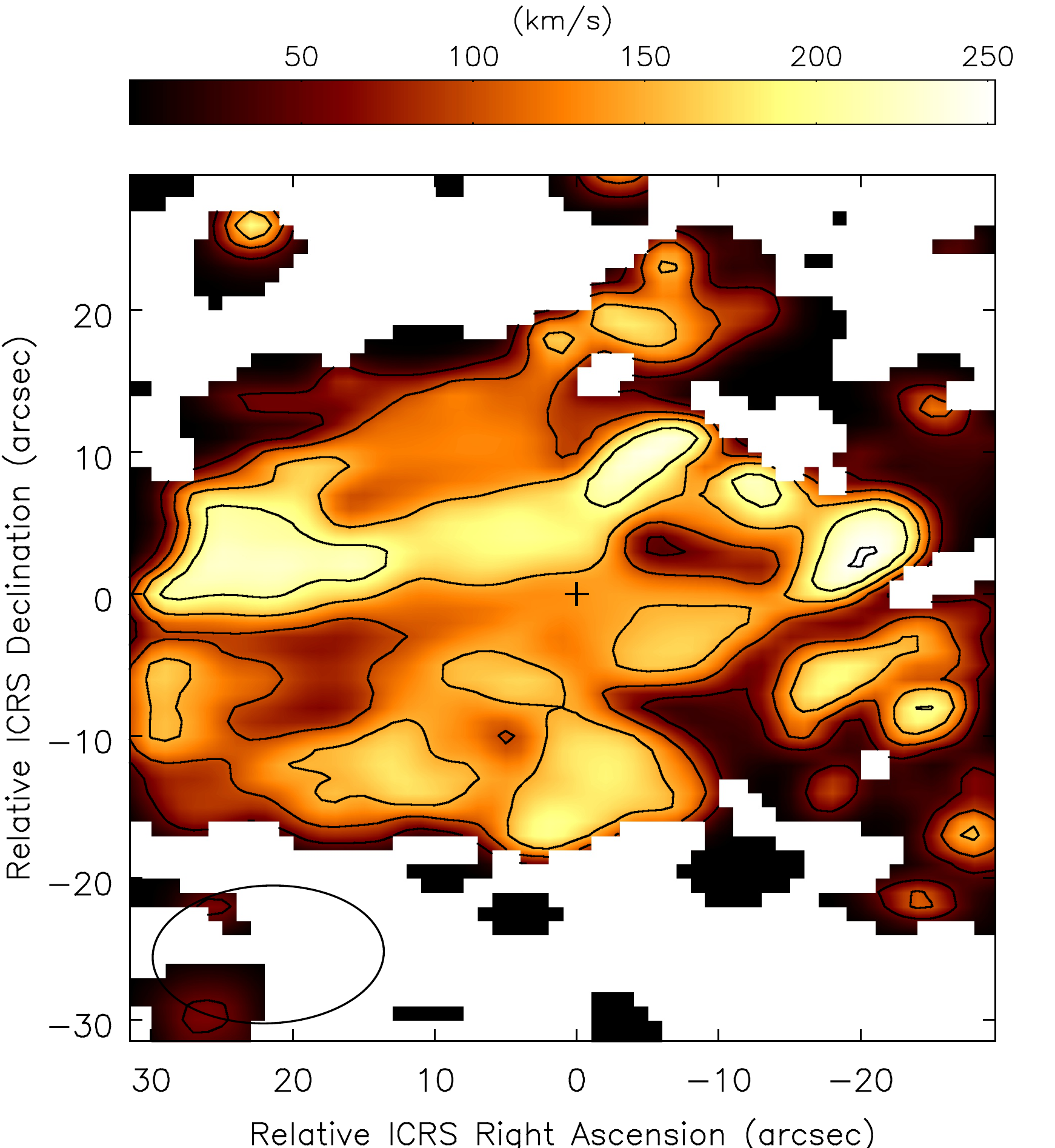}
        \caption{ACA CO(3-2) moment maps, produced by selecting the velocity range to $-300<v[\rm km~s^{-1}]<300$. 
                The left panel shows the zeroth moment map (flux), with contours at [0.2, 0.3, 0.4, 0.5, 0.6]~Jy~beam$^{-1}$~km~s$^{-1}$; the middle panel shows the first moment map (velocity), with contours corresponding to [-250, -150, -50, 50, 150, 250]~km~s$^{-1}$; the right panel shows the second moment map ($\sigma_v$), with contours at [50, 100, 150, 200, 250]~km~s$^{-1}$. The black cross indicates the AGN position, whose coordinates are reported in Table~\ref{table:properties}.
        }\label{fig:co32_aca_moment_maps}
 \end{figure*}
 
 Figure~\ref{fig:co32_aca_moment_maps} shows the ACA CO(3-2) moment maps. These were produced with the task \texttt{immoments} in CASA by selecting the velocity range $v\in[-300, 300]$~\kms and masking pixels below a flux threshold of 0.04~Jy~beam$^{-1}$~km~s$^{-1}$. Since it was produced by applying a flux and a velocity threshold, the moment~0 map (left panel of Fig.~\ref{fig:co32_aca_moment_maps}) is biased against faint and high-velocity emission, and it is shown only for completeness. We refer to Fig~\ref{fig:co32_aca_maps} for an unbiased, total CO(3-2) flux map. 
 
The velocity map (middle panel of Fig.~\ref{fig:co32_aca_moment_maps}) does not display a regular velocity gradient. However, about $15\arcsec$ north-east of the AGN position, it is possible to identify an extended region of predominantly redshifted emission with $150\lesssim v[\rm km~s^{-1}]<200$, and an average velocity dispersion of $\sigma_v\sim 120$~\kms. Almost symmetrically, $\sim15\arcsec$ south-west of the AGN, the CO(3-2) is mostly blueshifted, with $-250<v[\rm km~s^{-1}]\lesssim-200$ and an overall low velocity dispersion of $\sigma_v< 50$~\kms.
In the central portion of the field, just south-west of the redshifted area, the moment~2 map shows a stripe of high-$\sigma_v$ CO emission extending towards the east, with $\sigma_v\gtrsim 200$~\kms.
Other clear distinctive features of the moment maps are two spots of redshifted CO emission about $20\arcsec$ west of the AGN, offset in declination by $\Delta\delta\sim +10\arcsec$ and $\Delta\delta\sim -10\arcsec$, respectively, and two regions of high $\sigma_v\gtrsim200$~\kms~between them.

Interestingly, the velocity dispersion peaks are not centred on cid\_346. This may favour accreting streams over the AGN-outflow interpretation for explaining the kinematics of the molecular CGM, since in the outflow scenario one would expect the $\sigma_v$ peak to coincide with the AGN position. However, we caution against an over-intepretation of Fig.~\ref{fig:co32_aca_moment_maps} since the very large beam of ACA would dilute any kinematic signature of kiloparsec-scale outflows. In fact, the ACA resolution of 80~kpc is much larger than any ISM structure (disk, ring, or outflow) ever observed in molecular tracers. At this resolution, even the two largest detections of molecular CGM reservoirs prior to this work, that is the Spiderweb protocluster (70~kpc, \citealt{Emonts+16}) and Candels-5001 (40~kpc, \citealt{Ginolfi+17}), would be unresolved. Therefore, the kinematic features resolved by ACA around cid\_346, shown in Figure~\ref{fig:co32_aca_moment_maps}, cannot be compared to anything known. Due to the poor resolution and low S/N of the data, none of these features can be studied in detail.

\end{appendix}

\end{document}